\documentclass{PoS}

\usepackage{graphicx}
\usepackage{amsmath}
\usepackage{upgreek}
\usepackage{color}

\title{Properties of non-local wave function equivalent potential with generalized derivative expansion}

\ShortTitle{Properties of wave function equivalent potential with generalized derivative expansion}

\author{\speaker{Takuya Sugiura}$^a$, 
        Keiko Murano$^a$, 
        Noriyoshi Ishii$^a$, and
        Makoto Oka$^{bc}$ \\
        \llap{$^a$}Research Center for Nuclear Physics(RCNP), Osaka University,
        Osaka 567-0047, Japan \\
        \llap{$^b$}Department of Physics, Tokyo Institute of Technology,
        Tokyo 152-8551, Japan \\ 
        \llap{$^c$}
        Advanced Science Research Center, Japan Atomic Energy Agency, 
        Tokai, Ibaraki, 319-1195, Japan \\
       E-mail: \email{sugiura@rcnp.osaka-u.ac.jp}}

\abstract{

We examine the properties of the wave-function-equivalent potentials
which HAL QCD collaboration has introduced.
We generalize the derivative expansion, and then apply it to
energy-independent and non-local potentials in a coupled-channel
model.
We observe that the expansion converges by comparing the scattering
phase shifts computed from these potentials with the exact one.
We have also found that the convergence can be improved by either
varying the choice of interpolating fields or tuning the expansion
scale in the generalized derivative expansion.
The results will be utilized in future lattice QCD calculations,
allowing for further application of the HAL QCD method.

}

\FullConference{34th annual International Symposium on Lattice Field Theory\\
		24-30 July 2016\\
		University of Southampton, UK}

\begin{document}

\vspace{-1.0em}
\section{Introduction}
\vspace{-0.7em}

Deriving the nuclear force from QCD is one of the most important
subjects in nuclear physics, while it has been a challenge because of
the non-perturbative nature of low-energy QCD.
Recently, a method to determine the nuclear force through lattice QCD
calculations has been introduced and vigorously developed by HAL QCD
collaboration~\cite{HAL-PRL2007,HAL-PTP2010}.
The method is based on the use of the 
{\em Nambu-Bethe-Salpeter (NBS) wave functions}
as inputs to the Schr\"odinger equation, so that the nuclear force is
determined as a non-local 
\footnote{ In this paper, we use the term {\em non-local} for the
  potentials that cannot be parameterized as $\langle x |\hat{V}|x'
  \rangle = V(x)\delta(x-x')$ in the coordinate space.  }
and energy-independent potential which reproduces all the NBS wave
functions simultaneously.
The non-locality of the HAL QCD potential is taken into account by the
derivative expansion, so that the potential is expressed as a power
series of spatial derivatives whose coefficients are determined from
the NBS wave functions for different energies.
In a previous study~\cite{Murano-PTP2011}, the authors studied the
validity of the derivative expansion in an indirect method: they
compared two local lowest-order HAL QCD potentials computed from the
NBS wave functions for c.m. energies $E\sim 0$~MeV and $E\sim 45$~MeV.
The two states appeared as the ground states with the periodic and the
anti-periodic spatial boundary conditions.  They found that the
potentials agreed within statistical errors.
However, it is desirable to evaluate the derivative expansion more
explicitly, since the higher-order terms might not be negligible for
higher-energy nucleon-nucleon scattering.

We perform a model calculation to examine the properties of the HAL
QCD potentials when the higher-order terms of the derivative expansion
are explicitly taken into account.
%
%
We also study the possibility of improving the convergence by
either varying the choice of interpolating fields or introducing a
non-locality scale parameter to the expansion.

\vspace{-0.7em}
\section{Model}
\vspace{-0.7em}

We employ a 1+1 dimensional coupled-channel model originally proposed
by M.~Birse \cite{Birse}.
We first construct a second-quantized non-relativistic Hamiltonian
$\hat{H}$, involving three different scalar fields, $\hat{p}(x)$,
$\hat{n}(x)$, and $\hat{n}'(x)$, which are analogous to proton,
neutron, and the first excited state of neutron with excitation energy
$\Delta$, respectively.
We consider an attractive square-well interaction for the pn-pn
coupling and a contact interaction for pn-pn$'$, while the pn$'$-pn$'$
interaction is assumed to be absent for simplicity.
To discuss the two-body scattering problem in this system, we consider
the two-particle states of pn and pn$'$.
We define the wave functions for these channels as
\begin{equation}
  \psi_0(x) \equiv \langle 0 | \hat{p}(x+y) \hat{n}(y) | \Psi (E)\rangle,
  \hspace{5pt}
  \psi_1(x) \equiv \langle 0 | \hat{p}(x+y) \hat{n}'(y) | \Psi (E)\rangle,
\end{equation}
where $|\Psi(E)\rangle$ denotes the energy eigenstate in the center of
mass frame satisfying $\hat{H}|\Psi (E)\rangle = E|\Psi (E)\rangle$.
We then find the wave functions satisfy the following coupled-channel
equations:
\begin{align}
\begin{split}
  \left[-\frac{1}{M}\frac{d^2}{dx^2}+V_0(x)-E\right]\psi_0(x) + 2g\delta(x)\psi_1(x)
  &= 0, \\
  \left[-\frac{1}{M}\frac{d^2}{dx^2}+\Delta-E\right]\psi_1(x) + 2g\delta(x)\psi_0(x)
  &= 0,
\end{split}
\label{eq:CCeq}
\end{align}
with
\begin{equation}
  V_0(x) = 
  \begin{cases}
    -V_0 & \mbox{for}\hspace{3mm} |x|<R  \\
    0    & \mbox{for}\hspace{3mm} |x|>R. 
  \end{cases}
\label{eq:square-well}
\end{equation}
The even-parity solution of Eqs.~\eqref{eq:CCeq} is obtained
analytically, which is conveniently utilized to implement the
derivative expansion with precision.  We exclusively focus on the
energy region $E<\Delta$ so that $\psi_1(x)$ vanishes at large
distance.

In this model, a general interpolating field for ``neutron'' is 
achieved as a linear combination of the fields $n(x)$ and $n'(x)$.
We define such a field according to
\begin{equation}
  \hat{\phi}_q(x) \equiv \hat{n}(x) + q\hat{n}'(x),
\end{equation}
where a real parameter $q$ is introduced to arrange the mixing of the
states. We refer to $q$ as the {\em field admixture parameter} in
this paper.
%
%
The NBS wave function with the fields $\hat{p}(x)$ and
$\hat{\phi}_q(x)$ is expressed as a linear combination of $\psi_0(x)$
and $\psi_1(x)$:
\begin{equation}
  \label{eq:Psi_q(x)}
  \Psi_q(x) 
  \equiv \langle 0 | \hat{p}(x+y) \hat{\phi}_q(y) | \Psi(E) \rangle
  = \psi_0(x) + q\psi_1(x).
\end{equation}
We utilize $\Psi_q(x)$ to construct HAL QCD potentials, and the
dependence on the choice of interpolating fields is studied by varying
the field admixture parameter $q$.

The contribution of $\psi_1(x)$ is suppressed at large distance, so
that the long-distance asymptotic behavior of our NBS wave function
reads
\begin{equation}
  \label{eq:as-behavior}
  \Psi_q(x)\to \psi_0(x)
  \simeq
  A \cos\left(k|x| + \delta(k)\right),
\end{equation}
where $\delta(k)$ denotes the scattering phase shift and $k\equiv
\sqrt{ME}$. Similar relation for the NBS wave function on the lattice
has been derived \cite{HAL-PTP2010} through the LSZ reduction formula.
This relation ensures that the HAL QCD potentials are faithful to the
scattering phase shift by construction.

\section{Non-Local Potential and Derivative Expansion}

Now we discuss how to construct a potential by using the NBS wave
functions~\eqref{eq:Psi_q(x)} for various different energies.
We start from the following Schr\"odinger equation in a finite box:
\begin{equation}
  \label{eq:Schroedinger}
  \left(-H_0+E_m\right)\Psi(x;E_m)=\int dx' V(x,x') \Psi(x';E_m),
\end{equation}
where $H_0\equiv -\frac{1}{M}\frac{d^2}{dx^2}$ is the free Hamiltonian
and $E_m$ are the discrete energy eigenvalues.  In general, the
potential has to be non-local to reproduce a given set of NBS wave
functions $\Psi(x;E_m)$ for different energies. The non-locality can
be taken into account by the (na\"ive) derivative expansion:
\begin{equation}
  \label{eq:naiveDE}
  V(x,x') = \sum_{n=0}^\infty u_n(x) 
  \left(\frac{\partial}{\partial x}\right)^n \delta(x-x'),
\end{equation}
where the derivatives in the higher order terms yield the
non-locality.

The na\"ive derivative expansion~\eqref{eq:naiveDE} does not work well
in the Birse model. 
With the na\"ive expansion, Eq.~\eqref{eq:Schroedinger} is rewritten as
\begin{equation}
  (-H_0+E_m)\Psi(x;E_m)=
  \sum_{n=0}^\infty u_n(x) \frac{d^n \Psi}{dx^n} (x;E_m).
\end{equation}
We find that $\frac{d^n\Psi}{dx^n}$ on the right-hand side are
singular at $x=0$ and $x=\pm R$, because of the $\delta(x)$ coupling
and the square-well edges in Eqs.~\eqref{eq:CCeq}, respectively.
To avoid the singularity, we introduce a generalized derivative
expansion as follows:
\begin{equation}
  \label{eq:generalizedDE}
  V(x,x')=
  \sum_{n=0}^\infty 
  v_n(x) \left(\frac{\partial}{\partial x}\right)^n \delta_\rho(x-x'),
  \hspace{2em}
  \delta_\rho(x-x') \equiv 
  \frac{\exp\left\{-(x-x')^2/\rho^2\right\}}{\sqrt{\pi}\rho},
\end{equation}
where $\rho$ is an arbitrary real parameter. In the following, we
refer to $\rho$ as the {\em Gaussian expansion scale}.
In the case of the generalized expansion, Eq.~\eqref{eq:Schroedinger}
is rewritten as
\begin{equation}
  (-H_0+E_m)\Psi(x;E_m)=
  \sum_{n=0}^\infty v_n(x) \frac{d^n \Phi_\rho}{dx^n} (x;E_m),
\end{equation}
where the smeared wave function 
$\Phi_\rho(x;E_m) \equiv \int dx'\, \delta_\rho(x-x') \Psi(x';E_m)$
appearing on the right-hand side is smooth everywhere so that it
causes no singularity.
In Ref.~\cite{Sugiura}, we give a proof that the generalized
derivative expansion is exact.  It is a natural generalization of the
na\"ive derivative expansion, since $\delta_\rho(x-x')$ is reduced to
$\delta(x-x')$ in the $\rho\to 0$ limit.
Moreover, we expect that the convergence of the generalized expansion
can be improved by tuning the expansion scale $\rho$.
We have made a few non-essential and technical modifications to
Eq.~\eqref{eq:generalizedDE} in presenting the results below. The
details are found in Ref.~\cite{Sugiura}.

In order to examine the convergence, we first truncate the summation
over $n$ in Eq.~\eqref{eq:generalizedDE} at a finite order, $N$.  The
coefficient functions $v_n(x)$ ($n=0,\cdots,N$) are then determined
from the NBS wave functions~\eqref{eq:Psi_q(x)} for the $(N+1)$
lowest-lying energy levels in a finite box.
Because of the $\delta(x)$ coupling terms in Eq.~\eqref{eq:CCeq}, each
coefficient $v_n(x)$ shall be decomposed into a regular part
$\tilde{v}_n(x)$ and the $\delta$ functional singular part as
$v_n(x)=\tilde{v}_n(x)+g_n\delta(x)$.  The singular part is treated
separately to determine the strength constants $g_n$ from the same NBS
wave functions.
We then solve the Lippmann-Schwinger equation with these HAL QCD
potentials to calculate the scattering phase shift defined as
Eq.~\eqref{eq:as-behavior}.  The result is compared to the analytic
solution of the model to discuss the convergence of the expansion.

\section{Results}

We employ the parameter set given in Ref.~\cite{Birse}, i.e.
$MV_0=1/R^2$, $M\Delta=6/R^2$, $Mg=6/R$, and we take $R=1$ and $M=1$.
The solution of coupled-channel Eqs.~\eqref{eq:CCeq} is analytically
obtained under the twisted boundary condition (TBC)
\begin{align}
\begin{split}
  \psi_{0,1}(x+2L) &= e^{i\theta}\psi_{0,1}(x), \\
  \psi_{0,1}^*(x)  &= \psi_{0,1}(-x),
\end{split}
\end{align}
with $L=10$ and $\theta=\pi/2$.
The second condition implies that the real part of the solution is
parity-even and the imaginary part is parity-odd.  Thus we only use
the real part to consider the parity-even sector.
The application of the commonly used (anti-)periodic BC leads to a
technical problem around the boundary, since the even-order (odd-order)
derivatives of the NBS wave functions would be zero for all energies.
With this parameter set and the TBC, we find a bound state at
$E=E_0\simeq-33.7$ and 15 excited states, $E=E_1,\cdots,E_{15}$ below
the threshold $E=\Delta$.  A HAL QCD potential with truncation order
$N$ is determined through the Schr\"odinger equations for
$E=E_0,\cdots,E_N$.

\begin{figure}[h]
  \includegraphics[width=.5\textwidth,bb=50 70 410 282]{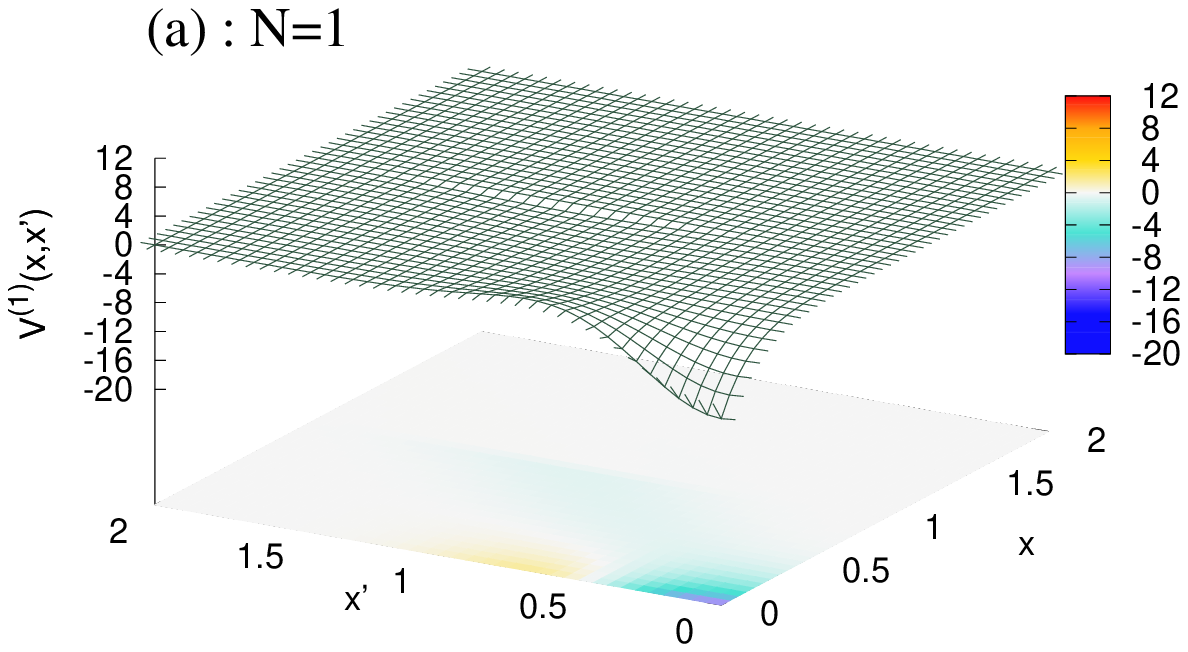}
  \includegraphics[width=.5\textwidth,bb=50 70 410 282]{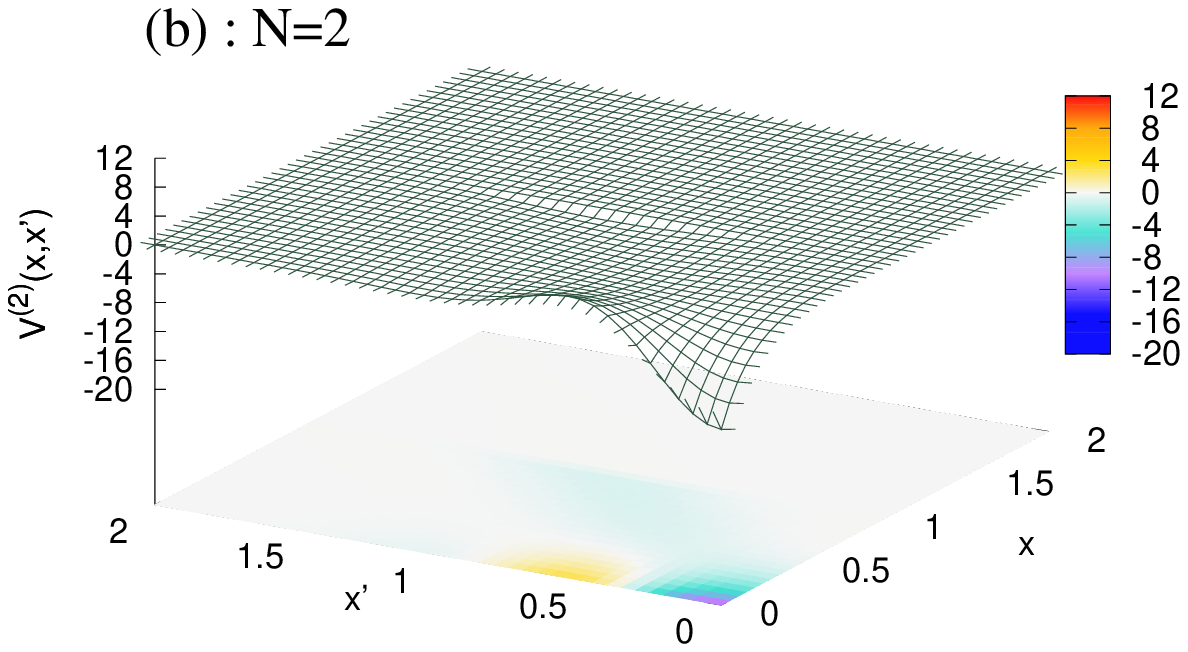}
  \includegraphics[width=.5\textwidth,bb=50 70 410 282]{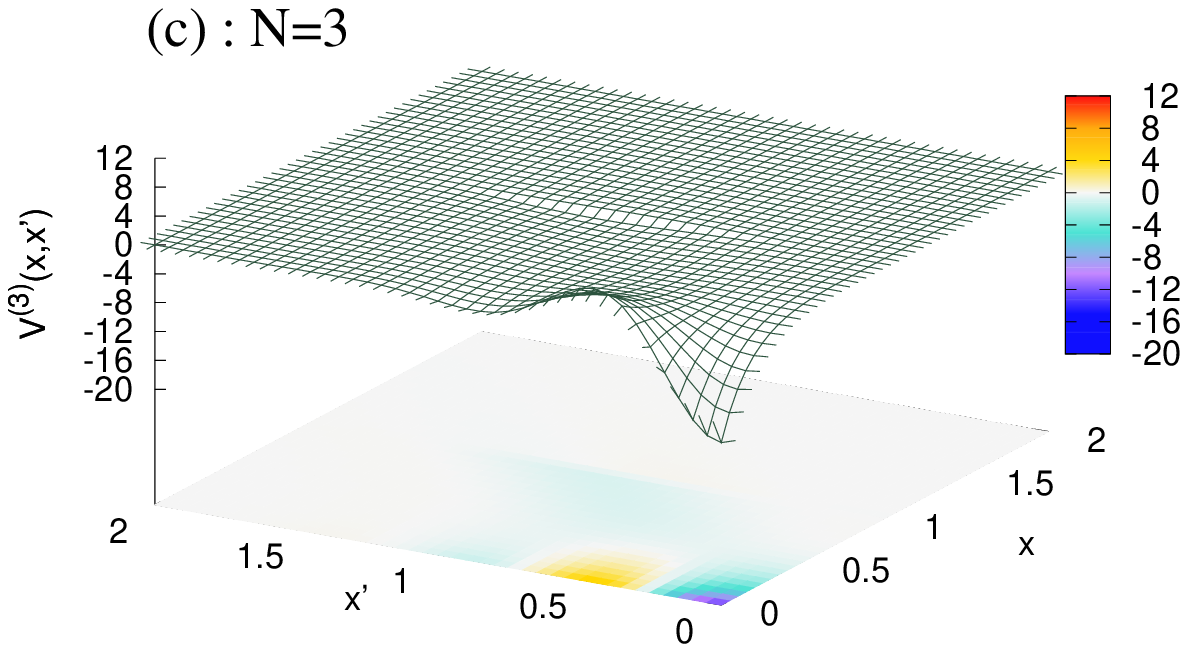}
  \includegraphics[width=.5\textwidth,bb=50 70 410 282]{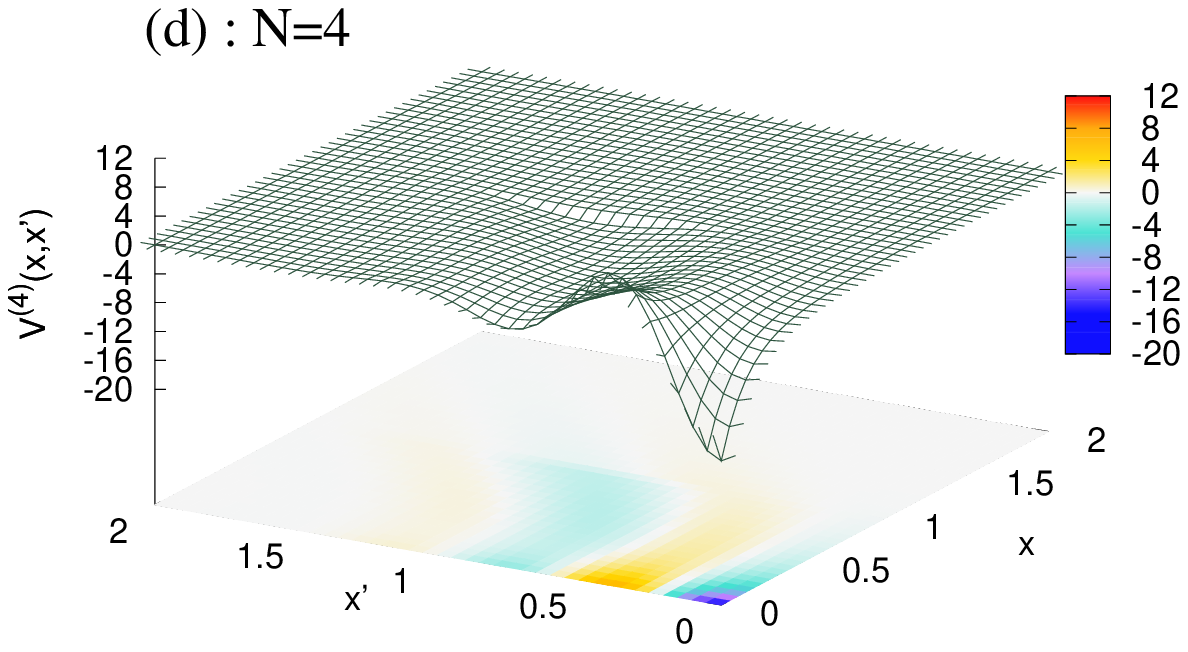}
  \includegraphics[width=.5\textwidth,bb=50 70 410 282]{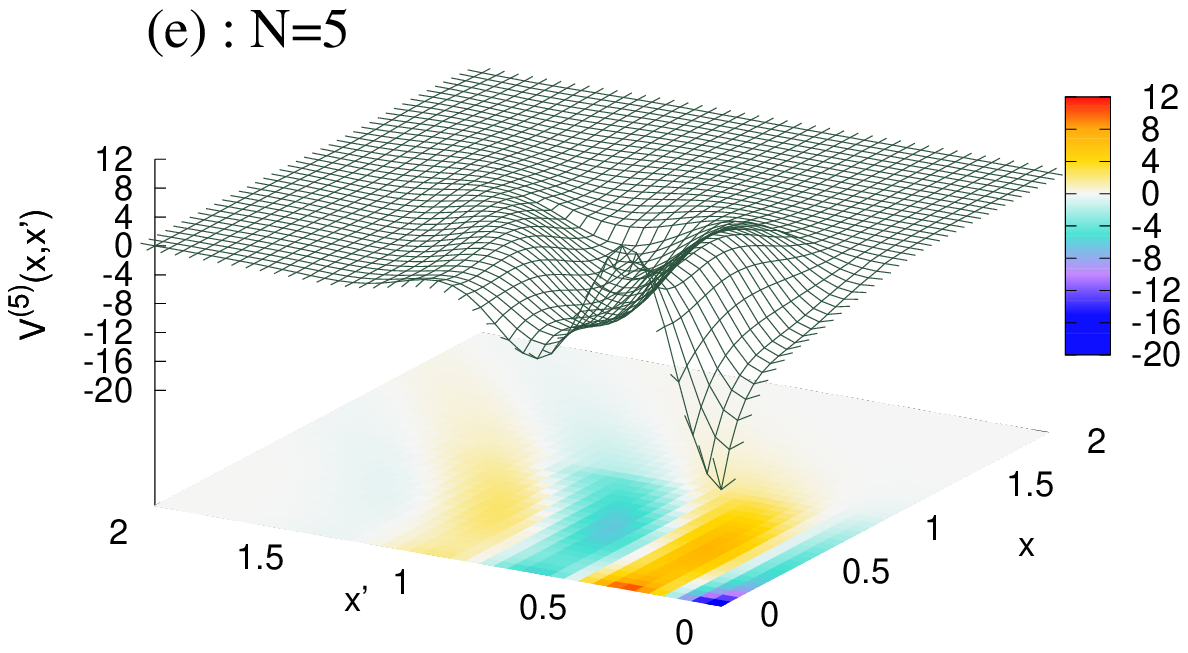}
  \caption{ 
    \label{fig:NLpots}
    Non-local potentials $V(x,x')$ with truncation orders of (a)
    $N=1$, (b) $N=2$, (c) $N=3$, (d) $N=4$, and (e) $N=5$.  The field
    adxmixture parameter and the Gaussian expansion scale are fixed to
    $q=+0.2$ and $\rho=0.5$, respectively.
  }
\end{figure}

In Fig.~\ref{fig:NLpots}, we show the regular parts of the HAL QCD
potentials with truncation orders $N=1,2,3,4,5$ and fixed $q=+0.2$ and
$\rho=0.5$.
We see that the structure of the potentials varies depending on the
truncation order, as the magnitude tends to become larger for larger
$N$.
%
%
All the potentials are finite-ranged, and we find $V(x,x')\simeq 0$
for $|x|>3$; therefore, we safely regard $V(x,x')=0$ for $|x|>4$ in
calculating the phase shift.

\begin{figure}[h]
  \centering
  \includegraphics[width=.5\textwidth]{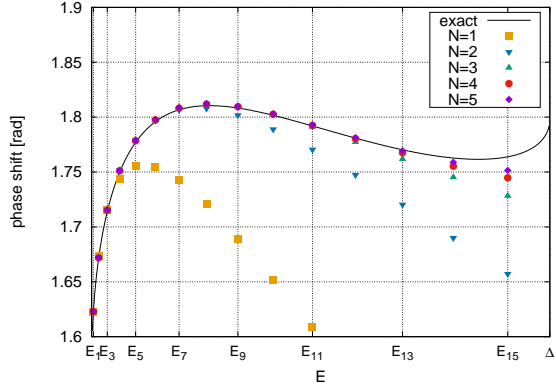}
  \caption{
    \label{fig:ps-N}
    Scattering phase shifts $\delta_N(E)$ at truncations
    $N=1,2,3,4,5$, compared to the exact one $\delta_{exact}(E)$
    extracted from the analytic solution of the model.
    Here $q=+0.2$ and $\rho=0.5$ are fixed.
  }
\end{figure}

Figure~\ref{fig:ps-N} shows the scattering phase shifts $\delta_N(E)$
computed from the HAL QCD potentials in Fig.~\ref{fig:NLpots}, i.e.,
the ones with fixed $q=+0.2$ and $\rho=0.5$ and varying truncation
orders $N=1,\cdots,5$.  The exact phase shift $\delta_{exact}(E)$
extracted from the analytic solution is also plotted for comparison.
We see that each of $\delta_N$ reproduces $\delta_{exact}$ in a small
energy region, but deviates away for the higher energies.  However,
the deviation becomes smaller as $N$ increases and more terms are
included.  It indicates that the generalized derivative expansion
actually converges to give the correct phase shift.

The agreement between $\delta_N$ and $\delta_{exact}$ for $E\leq E_N$
is ensured by construction, since the NBS wave functions for these
energies are used as inputs to the Schr\"odinger equation.
The result also shows that the agreement extends beyond these
energies, indicating that extrapolation to higher energies is
possible.
In the following, we refer to such a region with nontrivial agreement
as the {\em extrapolation region}, and use it to discuss the
convergence.

Note that our result in the 1+1 dimensional space-time shows the
relation $\delta(E=0)=\pi/2$ with the existence of a single bound
state, which is contrasted to
$\delta_l(E=0)-\delta_l(E=\infty)=n_l\pi$ in 1+3 dimension with $n_l$
bound states of angular momentum $l$.

\begin{figure}[h]
\begin{tabular}{cc}
  \begin{minipage}{0.49\hsize}
  \centering
  \includegraphics[width=\textwidth]{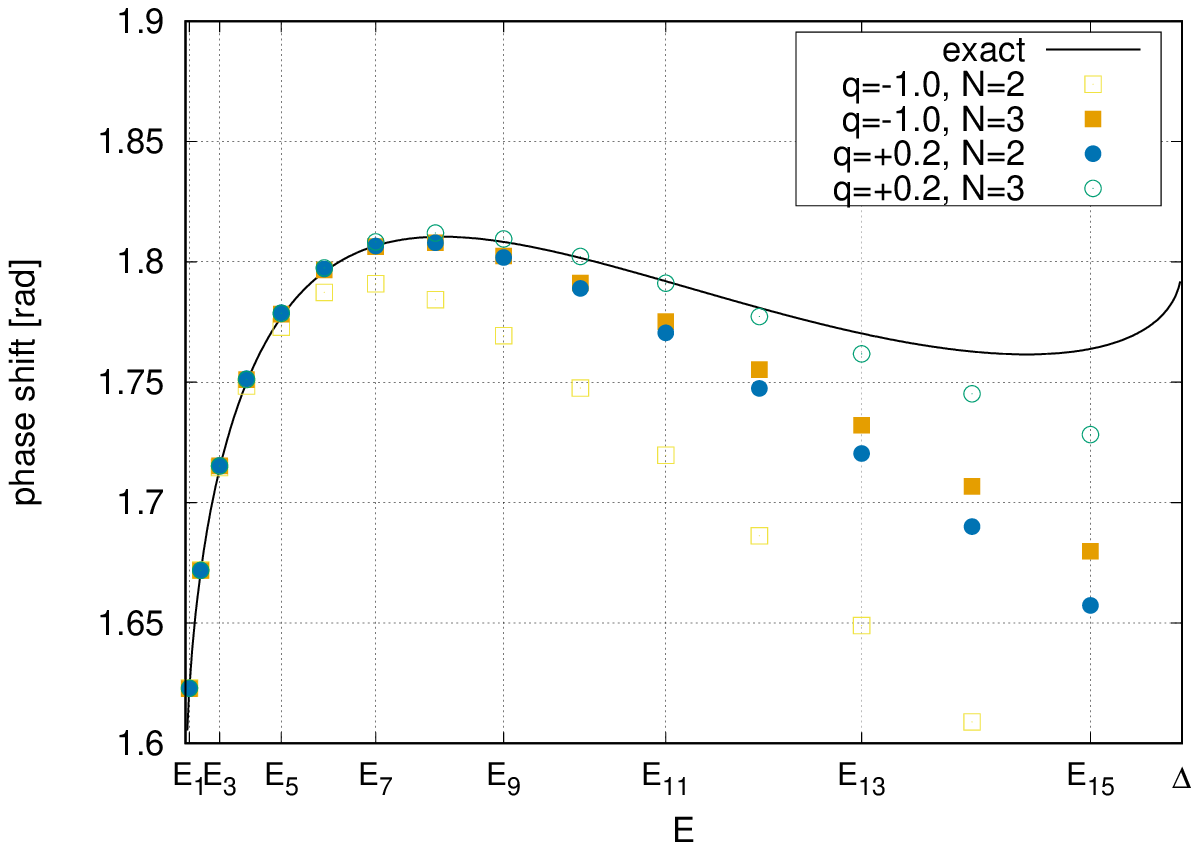}
  \caption{
    \label{fig:ps-q}
    Phase shifts with $q=-1.0,+0.2$ and
    fixed $\rho=0.5$, at $N=2,3$.
  }
  \end{minipage}
  \hspace{0.02\hsize}
  \begin{minipage}{0.49\hsize}
  \centering
  \includegraphics[width=\textwidth]{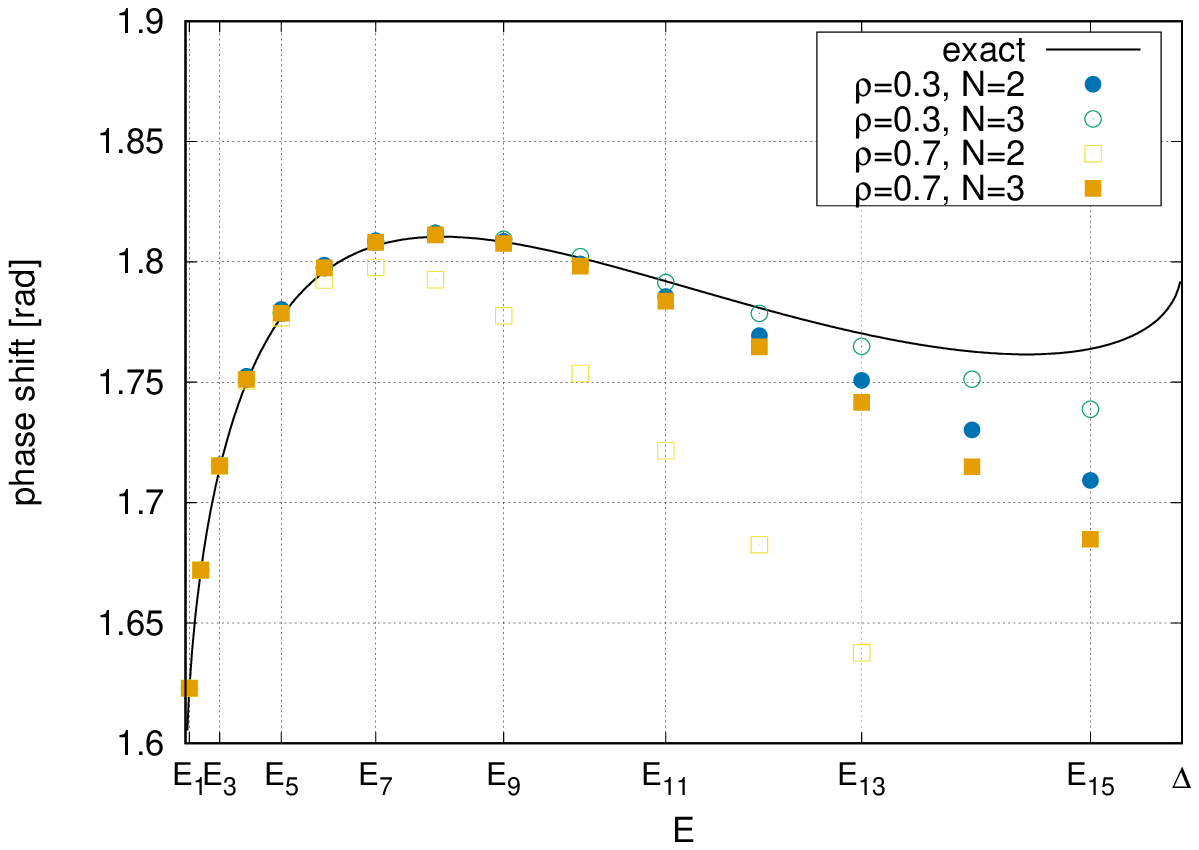}
  \caption{
    \label{fig:ps-rho}
    Phase shifts with fixed $q=+0.2$ and
    $\rho=0.3, 0.7$, at $N=2,3$.
  }
  \end{minipage}
\end{tabular}
\end{figure}

To study the dependence on the choice of interpolating fields, we vary
the field admixture parameter $q$. In Fig.~\ref{fig:ps-q}, we show the
phase shifts with $q=-1.0, +0.2$ and fixed $\rho=0.5$ at truncation
orders $N=2,3$.
We see that the derivative expansion is also convergent with $q=-1.0$,
since the result with $(q,N)=(-1.0, 3)$ shows better agreement with
the exact one than that of $(q,N)=(-1.0, 2)$.
Moreover, we see that the upper limit of the extrapolation region is
comparable between the $(q,N)=(-1.0, 3)$ and $(q,N)=(+0.2, 2)$ cases,
despite the different truncation orders.
It implies that the convergence can be improved by tuning the choice
of interpolating fields.

We expect the Gaussian expansion scale $\rho$ can also be used
to improve the convergence of the generalized derivative expansion.
In Fig.~\ref{fig:ps-rho}, we show the scattering phase shifts with
$\rho=0.3,0.5$ and $N=2, 3$, while the field admixture parameter is
fixed to $q=+0.2$.
The convergence of the generalized expansion is confirmed once again
for each choice of $\rho$.  The upper limit of the extrapolation
region in the case of $(\rho,N)=(0.3,2)$ is shown to be similar to
that of $(\rho,N)=(0.7,3)$, although the truncation order is smaller
and thus fewer number of NBS wave functions are used as inputs.
We therefore observe that we can improve the convergence by properly
choosing the expansion scale in the generalized derivative expansion,
as well as tuning interpolating fields.
In practice, this property will be of great use, since the generalized
derivative expansion is also applicable to lattice QCD calculations,
and varying the $\rho$ value does not require any additional
computational cost.
%

\section{Summary}

We have studied the properties of the non-local potentials introduced
by HAL QCD collaboration.  We have employed an analytically solvable
coupled-channel model to precisely perform the derivative expansion to
higher orders, since it would require huge computational cost in
lattice QCD.  The generalized derivative expansion has been introduced
to be applied to the present model, which involves unsmooth NBS wave
functions.
We have investigated the convergence of the generalized
expansion by evaluating the validity of the truncation at a finite
order.

We have observed that the generalized derivative expansion converges
such that the agreement between the phase shift obtained from the HAL
QCD potentials and the exact one becomes better for higher truncation
orders.  The result indicates that extrapolation to higher energies
than those used as input is possible.  We have also found that the
convergence can be improved by either varying the choice of hadron
interpolating fields or properly choosing the expansion scale which
appears in the generalized derivative expansion.

\vspace{1em}\noindent 
{\bf Acknowledgements}

We would like to thank K.~Hiranuma, A.~Hosaka, and the members of HAL
QCD collaboration for helpful and enlightening discussions about this
study.
This work was supported by JSPS KAKENHI Grand Numbers JP25400244 and
JP25247036, and by MEXT as ``Priority Issue on Post-K computer''
(Elucidation of the Fundamental Laws and Evolution of the Universe)
and JICFuS.

\end{document}